\documentclass[runningheads]{llncs}
\usepackage[T1]{fontenc}
\usepackage{graphicx}
\usepackage{hyperref}
\usepackage{makecell}
\usepackage{adjustbox}
\usepackage[misc]{ifsym}
\usepackage{color}

\begin{document}

\newcommand{\repeatthanks}{\textsuperscript{\thefootnote}}

\title{Contrastive Pretraining for Echocardiography Segmentation with Limited Data}
\titlerunning{Contrastive Pretraining for Echocardiography Segmentation}

\author{AUTHOR_NAME$^{(\textrm{\Letter})}$}
\author{Mohamed Saeed\thanks{Contributed equally}\orcidID{0000-0003-3222-7675}$^{(\textrm{\Letter})}$\and
Rand Muhtaseb\repeatthanks\orcidID{0000-0003-2604-5429} \and
Mohammad Yaqub\orcidID{0000-0001-6896-1105}}
\authorrunning{M. Saeed*, R. Muhtaseb* \& M. Yaqub}

\institute{Mohamed bin Zayed University of Artificial Intelligence, Abu Dhabi,\\United Arab Emirates\\
\email{\{mohamed.saeed,rand.muhtaseb,mohammad.yaqub\}@mbzuai.ac.ae}}
\maketitle              
\begin{abstract}
Contrastive learning has proven useful in many applications where access to labelled data is limited.  The lack of annotated data is particularly problematic in medical image segmentation as it is difficult to have clinical experts manually annotate large volumes of data such as cardiac structures in ultrasound images of the heart. In this paper, We propose a self supervised contrastive learning method to segment the left ventricle from echocardiography when limited annotated images exist.  Furthermore, we study the effect of contrastive pretraining on two well-known segmentation networks, UNet and DeepLabV3. Our results show that contrastive pretraining helps improve the performance on left ventricle segmentation, particularly when annotated data is scarce. We show how to  achieve comparable results to state-of-the-art fully supervised algorithms when we train our models in a self-supervised fashion followed by fine-tuning on just 5\% of the data. We show that our solution outperforms  what is currently published on a large public dataset (EchoNet-Dynamic) achieving a Dice score of 0.9252. We also compare the performance of our solution on another smaller dataset (CAMUS) to demonstrate the generalizability of our proposed solution. The code is available at (\url{https://github.com/BioMedIA-MBZUAI/contrastive-echo}).

\keywords{Contrastive Learning \and Segmentation \and Echocardiography \and Ultrasound \and SimCLR \and BYOL \and Self-supervised.}
\end{abstract}

\section{Introduction}

Echocardiography is a valuable diagnostic tool in cardiovascular disease as it can rapidly locate the presence of some abnormalities within the heart.  This involves the quantification of heart structures such as the left ventricle. However, there is a lot of room for error in this process due to factors such as operator variability, patient's characteristics e.g. high body mass index, or low image quality compared to other imaging modalities \cite{alsharqi2018artificial}.
 
Deep learning solutions can help automate the annotation process, but they are limited by the quantity and quality of labelled training data which can be difficult or expensive to obtain. For the problem of left ventricle segmentation in particular, previous works have had some success in detecting and quantifying heart abnormalities but there is room for improvement, potentially with the acquisition of more data \cite{kusunose2019utilization}. Since annotation of the raw data is cumbersome, self-supervised learning helps make use of unlabelled data that does not require input from clinical experts. In similar tasks such as view classification of echocardiography images, contrastive pretraining on unlabelled data showed impressive improvements in results \cite{chartsias2021contrastive}. This indicates that there is a potential benefit for segmentation problems given that the features learned for classification should not be too dissimilar.

\section{Related Work}
In this section, we aim to give a brief revisit to important concepts which our paper investigates. We believe this is important to make our work clearer to a wide range of audiences. 

\subsection{Segmentation Networks}

We investigate two well-known segmentation networks, UNet \cite{ronneberger2015u} and DeepLabV3 \cite{chen2017deeplab}, which have demonstrated huge success in many segmentation problems and this is why we have chosen them. UNet is a fully convolutional network that consists of a contracting path (encoder) and an expanding path (decoder) in a U-shaped architecture. Features are extracted by the contracting path and then upsampled gradually by the expanding path, with skip connections between corresponding layers in the contracting and expanding paths. The second network is DeepLabV3 which has initially shown great performance on semantic segmentation of natural images. It introduces an Atrous Spatial Pyramid Pooling (ASPP) module \cite{aspp} that utilizes atrous (dilated) convolutions at different rates to solve the problem of object scale variations in addition to expanding the receptive field while keeping the feature maps' spatial dimensions. ASPP consists of multiple dilated convolutions at different rates stacked in parallel followed by a concatenation of the outputs of said convolutions. Features from the encoder are passed through the ASPP module before upsampling back to the original resolution. In the following subsection, we review the use of these two networks in echocardiographic left ventricle segmentation.

\subsection{Ventricular Segmentation}

One example pertaining to the use of deep learning in ventricular segmentation made use of a UNet to segment the left ventricle in more than 1500 images from ultrasound videos of 100 patients \cite{smistad20172d}. The network was trained on the output of another segmentation algorithm that used a Kalman filter. Expert annotation was only available for 52 of the images, so the dataset was expanded by automatically annotating more examples using the Kalman filter based algorithm. Consequently, the UNet trained on this data was able to achieve a Dice score of 0.87, outperforming the previous algorithm.
 
Later work by \cite{moradi2019mfp} proposed a modification to the UNet architecture by combining it with a feature pyramid network. This was trained for left ventricle segmentation on the publicly available Cardiac Acquisitions for Multi-structure Ultrasound Segmentation (CAMUS) dataset \cite{leclerc2019deep} which consists of two- and four-chamber ultrasound images from 500 patients. Testing was then done on an external dataset of 137 four-chamber view images. Results showed that this architecture outperformed other state-of-the-art methods, achieving a Dice score of 0.953 on the test set.

Recent work proposed ResDUnet \cite{ResDUnet}, which is a new deep learning method based on UNet, that integrates cascaded dilated convolution to extract features at difference scales, and deploys an improved version of the residual blocks instead of the standard UNet blocks to ease the training task. ResDUnet outperforms state-of-the-art methods on the CAMUS dataset with a Dice score of 0.951 compared to 0.939 with a standard UNet.

Furthermore, \cite{ouyang2020video} attempted the same task, training on their large publicly available EchoNet-Dynamic dataset \cite{ouyang2019echonet}, containing 20,060 annotated images from 10,030 patients. A DeepLabV3 \cite{chen2017deeplab} network was chosen for this task and obtained a Dice score of 0.9211.

\subsection{Contrastive Learning}

Whilst there are multiple published contrastive learning algorithms in the literature, we have chosen to investigate two commonly used ones, namely SimCLR and BYOL.

\begin{figure}[h!]
    \centering
    \includegraphics[width=0.7\textwidth]{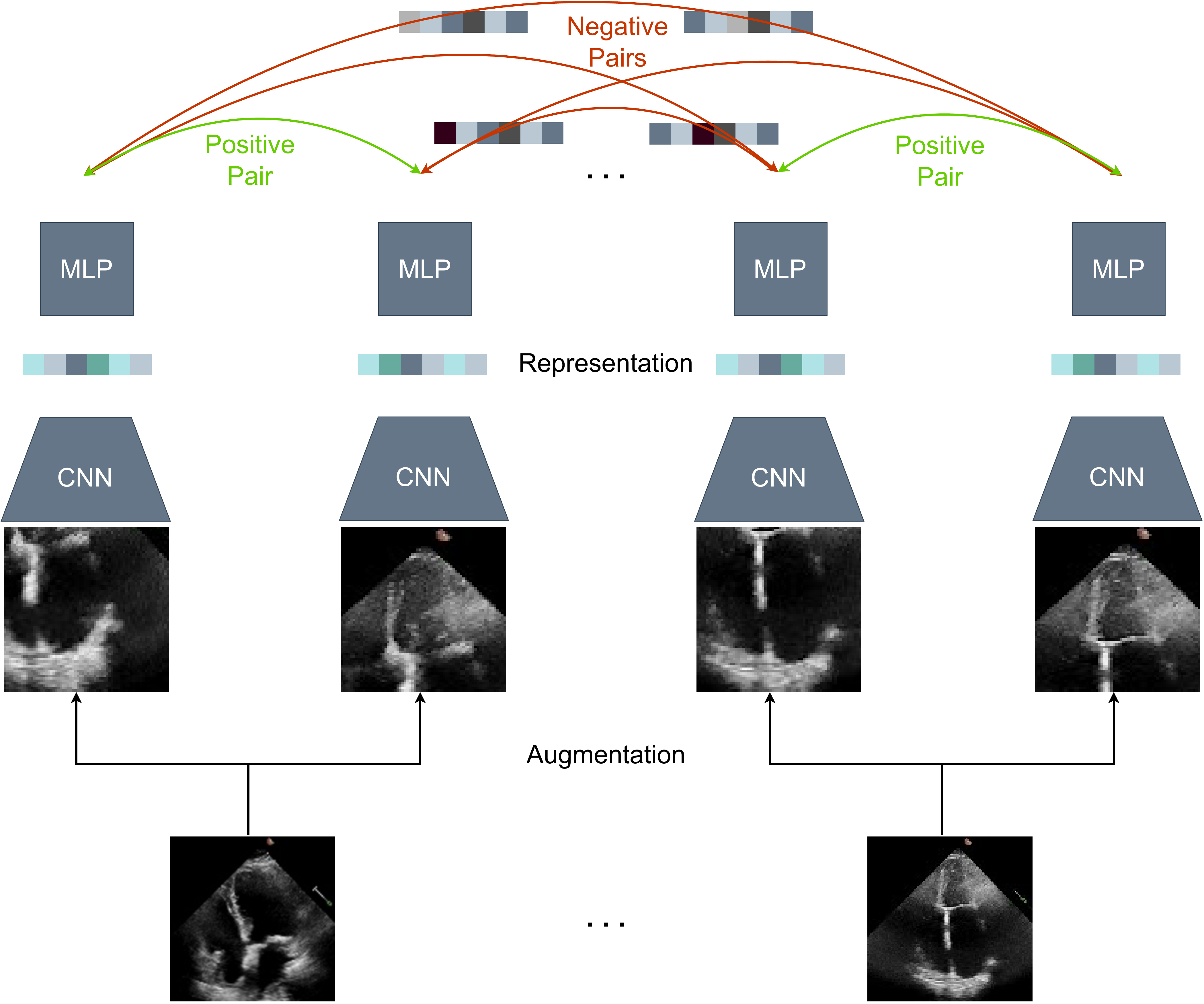}
    \caption{SimCLR framework applied to the echo images. The goal of the contrastive loss is to attract the positive pairs by maximizing their agreement, while repelling the negative pairs by maximizing their disagreement.}
    \label{fig:archs}
\end{figure}

\subsubsection{SimCLR}

SimCLR \cite{chen2020simple} is a popular framework for self-supervised contrastive learning, used to learn representations from unlabelled data. In essence, SimCLR creates two augmented versions of every input image. For each minibatch, one pair of augmented images coming from the same original image is chosen as the positive pair. All other pairs coming from different input images are considered negative pairs. The aim then becomes to maximize the agreement within the positive pair while simultaneously maximizing the disagreement between the positive pair and all the negative pairs. The framework begins with a base encoder which is a typical feature extractor such as a ResNet-50 \cite{he2016deep}. A projection head is added on top of this to map the encoded representation to a new space in which a contrastive loss based on cosine similarity is applied. Figure \ref{fig:archs} gives an overall view of how SimCLR works.

\subsubsection{BYOL}

\begin{figure}[h!]
    \centering
    \includegraphics[width=\textwidth]{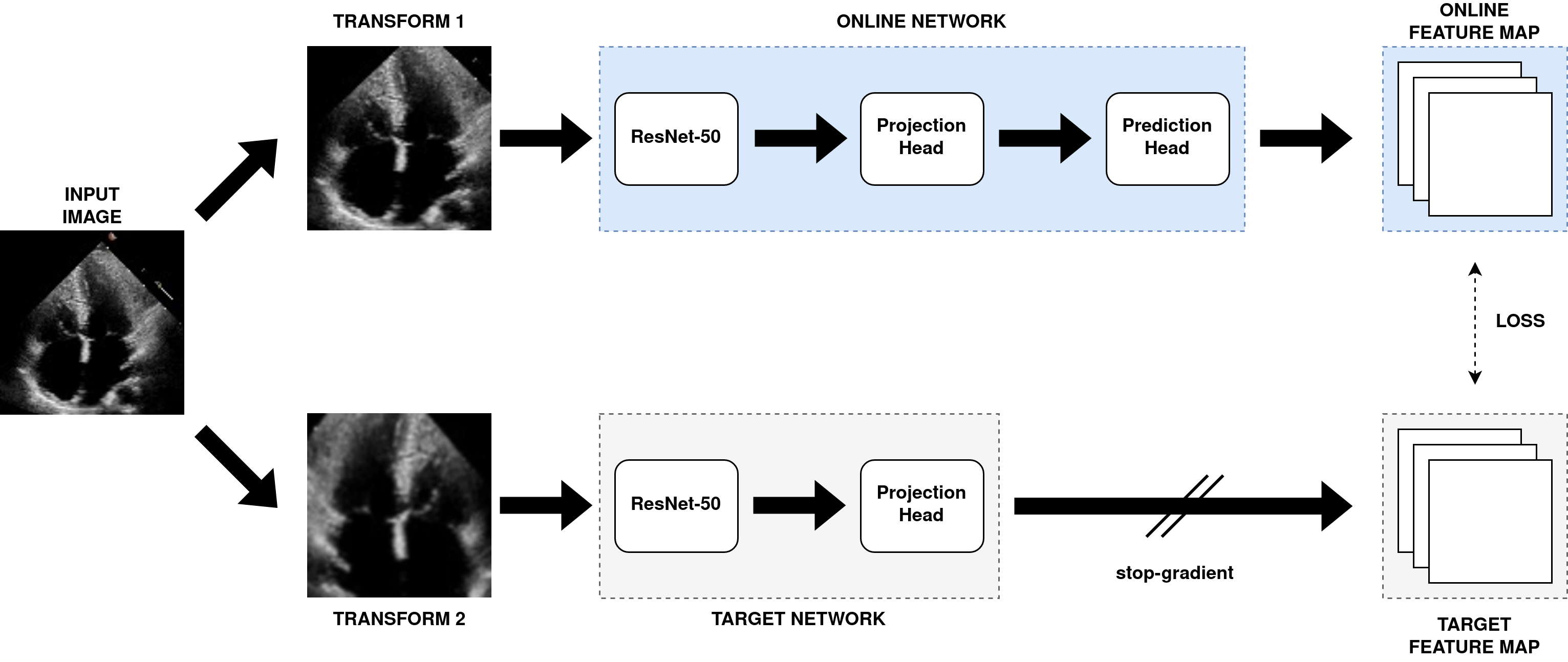}
    \caption{BYOL framework applied to the echo images. The aim is for the two networks to eventually produce a common representation given differently augmented versions of the same input image.}
    \label{fig:archb}
\end{figure}

Meanwhile, Bootstrap Your Own Latent (BYOL) \cite{grill2020bootstrap} uses a similar contrastive approach to SimCLR but without negative pairs, and for this reason we chose it to compare the effect of this difference on the contrastive pretraining. It always uses a single pair of images which are transformed versions of the same input. The framework allows representation learning by making use of two networks (called the \textit{online} and \textit{target} network). The online network is trained to predict the output of the target network. Meanwhile, the target network's weights are updated with just an exponential moving average of the online network. The two networks are mostly identical, having an encoder (usually a ResNet-50), followed by a projection head which linearly projects the encoder's features onto a different space. The only difference is that the online network has an added predictor head, which is simply another linear projection. Figure \ref{fig:archb} outlines the architecture. During training, the online network learns by attempting to maximize the agreement between the outputs from the two networks by minimizing a contrastive loss which simplifies to twice the negative of the cosine similarity between the two networks' outputs.

\section{Methods}

In this paper, we developed a solution to segment the left ventricle in echocardiographic images that is based on self-supervised contrastive learning. We argue why this could be a better approach than full supervision. This section describes the used data, the setup and the conducted experiments.

\subsection{Datasets}

\subsubsection{EchoNet-Dynamic}
The EchoNet-Dynamic dataset \cite{echonetdynamic} consists of 10,036 videos of apical four-chamber (A4C) view for patients who had an echocardiography between 2016 and 2018 at Stanford Health Care. Each video (encoded in RGB) consists of a sequence of 112 x 112 2D images extracted from the Digital Imaging and Communications In Medicine (DICOM) file and labeled with the corresponding left ventricle tracing, ejection fraction (EF), volume at end-systole (ES) and volume at end-diastole (ED) by expert sonographers. For each video, two frames (ES and ED) are annotated with manual segmentation. To the best of our knowledge this is currently the largest publicly available dataset for left ventricle segmentation, making it ideal for our contrastive task, given that it has a large amount of both labelled and unlabelled data.

\subsubsection{CAMUS}
The Cardiac Acquisitions for Multi-structure Ultrasound Segmentation (CAMUS) dataset \cite{leclerc2019deep} contains scans of 500 patients who underwent echocardiography at the University
Hospital of St Etienne in France. Each patient's data is labelled with the corresponding left ventricle ejection fraction (EF) and volumes at end-systole (ES) and end-diastole (ED). Annotations include tracings of the left ventricle endocardium, myocardium and the left atrium (LA) for both apical two-chamber (A2C) and apical four-chamber (A4C) views of the heart. Training and testing sets consist of 450 annotated and 50 unannotated videos, respectively. We found that 50 patients are missing from the training set, resulting in data of only 400 patients  for the training set. We have chosen this small dataset to investigate the importance of contrastive learning while experimenting with limited data.

\subsection{Experimental setup}

\begin{figure}[t!]
    \centering
    \includegraphics[width=\textwidth]{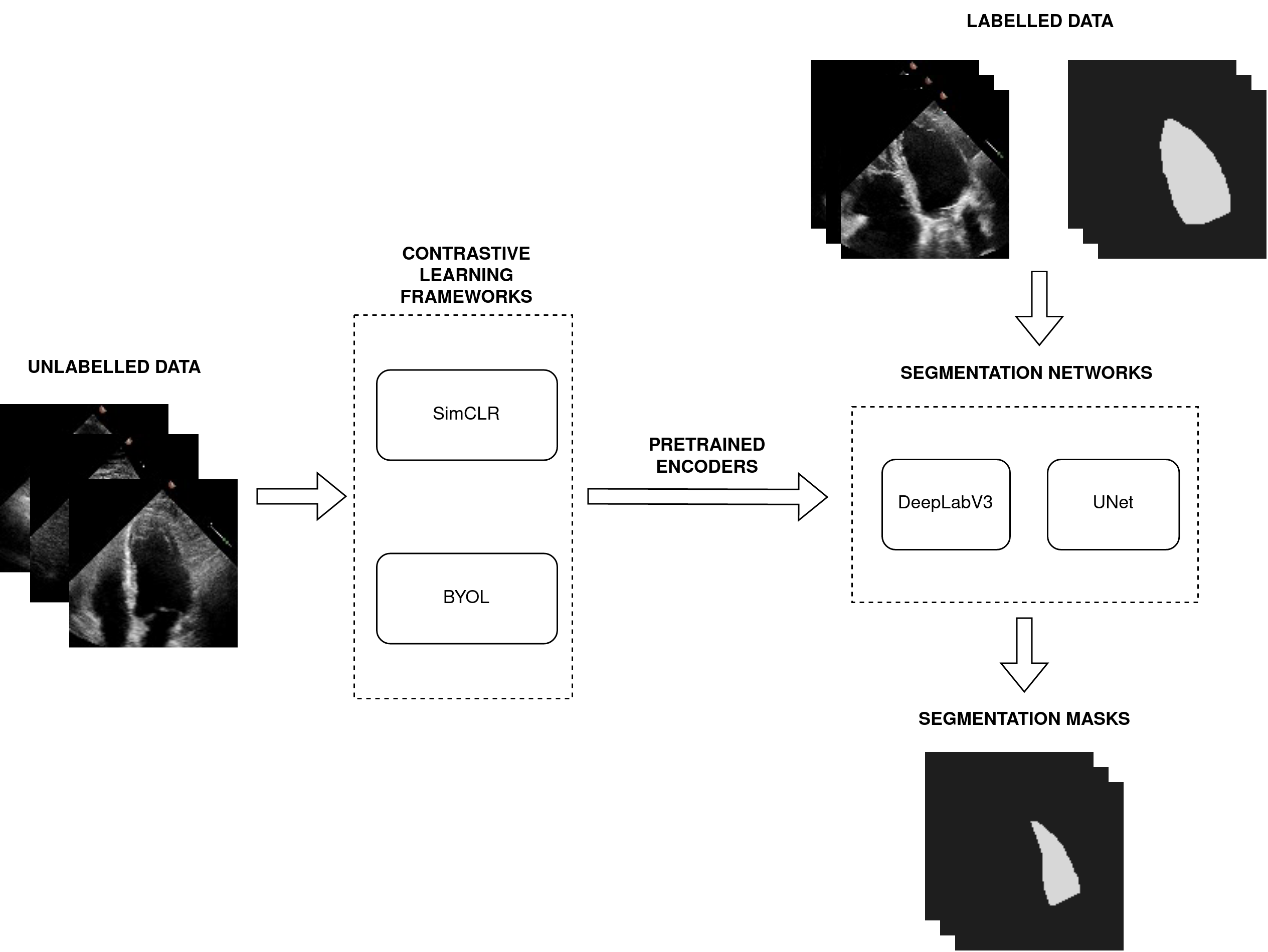}
    \caption{Overall experimental setup: The encoders are first trained on unlabelled data using SimCLR and BYOL. The pretrained weights are then loaded into DeepLabV3 and UNet to train on labelled data for the task of left ventricle segmentation.}
    \label{fig:arch}
\end{figure}

We experiment with SimCLR and BYOL pretraining \textit{(pretext task)} for left ventricle segmentation on the EchoNet-Dynamic and CAMUS datasets. First, we pretrained a DeepLabV3 backbone (ResNet-50 with atrous convolutions \cite{chen2017deeplab}) and a UNet backbone (original UNet encoder) with both SimCLR and BYOL. For the pretraining, unlabelled frames from the datasets are used. Thereafter, the pretrained backbones were used to train the segmentation networks, DeepLabV3 and UNet \textit{(downstream task)}. Figure \ref{fig:arch} outlines the overall experiment setup. The downstream segmentation experiments were done with 100\%, 50\% 25\% and 5\% of the available labelled data. In addition, we compare the SimCLR and BYOL pretrained backbones to randomly initialized and ImageNet pretrained (fully supervised) ones to see if self-supervision is beneficial. For evaluation, the Dice similarity coefficient (DSC) is used as a metric. 

\begin{equation}
    DSC = 2 * \frac{intersection}{intersection + union}
\end{equation}

All images were resized to 224x224 pixels for both the pretext and downstream tasks, since we keep the same setup used in SimCLR \cite{chen2020simple}. Bilinear interpolation was used for the input images and nearest neighbour interpolation was used for the masks.

\subsubsection{Pretext task} 

All backbones were pretrained for 300 epochs on two NVIDIA A6000 GPUs. DeepLabV3 backbones were trained with a batch size of 128 (64 per device) and UNet backbones were trained with a batch size of 256 (128 per device). An Adam optimizer was used for the pretraining with a learning rate of 1e-3 for SimCLR and 0.2 for BYOL. These were chosen experimentally. For both SimCLR and BYOL, we use the augmentation strategy proposed in the SimCLR paper to see if these contrastive learning algorithms work out of the box for ventricular segmentation. The augmentations consist of random resized cropping (scales=[0.08,1.0], aspect ratios=[0.75,1.33]), color distortions (jittering with brightness=0.8, contrast=0.8, saturation=0.8 and hue=0.2), Gaussian blurring (kernel size=23) and random horizontal flipping, following the default settings in \cite{chen2020simple}.

\subsubsection{Downstream task} 

The segmentation tasks were trained on a single A6000 GPU with a batch size of 128. A Madgrad \cite{madgrad} optimizer was used because it was found to converge better and faster than other optimizers and hyperparameters were selected experimentally. The base learning rate was 1e-4 for DeepLabV3 experiments and 1e-5 for UNet experiments.

\subsection{EchoNet-Dynamic Experiments}

The two annotated ES and ED frames from every video were used for the downstream task, resulting in 14,920 images for training, 2,576 for validation and 2,552 for testing. This is the same setup as the original EchoNet-Dynamic paper to allow a fair comparison and ensure that the sets are separated by patient. Meanwhile, for pretraining, the unlabelled frames in between ES and ED were used. One random frame between ES and ED was used for each patient. This was done to avoid having frames that are too similar to each other. As a result, the pretraining training set consisted of 7460 images, and the validation set contained 1288 images.

\subsection{CAMUS Experiments}

For the downstream task, the 400 available \textit{annotated} videos were split into 300 for training, 50 for validation and 50 for testing. Two frames (ES and ED) were taken from each video, leading to a training set of 600 images, a validation set of 100 images and a testing set of 100 images. For pretraining, a random frame (not including ES or ED frame) was taken from each of the 300 training videos. In addition, to create a validation set, a random frame from each of the videos in the \textit{unannotated} CAMUS test set was used. These are samples from the held out CAMUS test set that were not used anywhere else in our experiments. Overall, the pretraining task used 300 training images and 50 validation images.

\begin{table}[h]
  {\caption{Summary of experiments conducted on the EchoNet-Dynamic Dataset with different fractions of data for the downstream task.}\label{tab:echo}}%
  {\begin{adjustbox}{width=\columnwidth,center}
  \begin{tabular}{|l|l|l|l|l|l|l|}
  \Xhline{2\arrayrulewidth}
  \bfseries No. & \bfseries Pretraining & \bfseries Network & \bfseries Dice (SD) 100\% & \bfseries Dice (SD) 50\% & \bfseries Dice (SD) 25\% & \bfseries Dice (SD) 5\%\\
  \Xhline{2\arrayrulewidth}
  1 & - & DeepLabV3 & 0.9204 (0.0483) &  0.9164 (0.0490) &  0.9090 (0.0559) & 0.8920 (0.0718)\\
  \hline
  2 & ImageNet & DeepLabV3 & 0.9229 (0.0461) &  0.9175 (0.0605) & 0.9142 (0.0625) & 0.8968 (0.0778)\\
  \hline
  3 & SimCLR & DeepLabV3 & \textbf{0.9252} (0.0476) & \textbf{0.9242} (0.0452) & \textbf{0.9190} (0.0509) & \textbf{0.9125} (0.0548)\\
  \hline
  4 & BYOL & DeepLabV3 & 0.9209 (0.0636) & 0.9042 (0.1027) & 0.8938 (0.0906) & 0.8816 (0.1111)\\
  \hline
  5 & - & UNet & 0.9151 (0.0557)  &  0.9100 (0.0583) & 0.9046 (0.0675) & 0.8915 (0.0814)\\
  \hline
  6 & SimCLR & UNet & 0.9185 (0.0493) & 0.9157 (0.0571) & 0.9078 (0.0619) & 0.9048 (0.0653)\\
  \hline
  7 & BYOL & UNet & 0.9070 (0.0718) & 0.8959 (0.0775)  & 0.8768 (0.0997)  & 0.8318 (0.1284)\\
  \hline
  \end{tabular}
  \end{adjustbox}}
\end{table}

\begin{table}[h]
  {\caption{Summary of experiments conducted on the CAMUS Dataset with different fractions of data for the downstream task.}\label{tab:camus}}%
  {\begin{adjustbox}{width=\columnwidth,center}
  \begin{tabular}{|l|l|l|l|l|l|l|}
  \Xhline{2\arrayrulewidth}
  \bfseries No. & \bfseries Pretraining & \bfseries Network & \bfseries Dice (SD) 100\% & \bfseries Dice (SD) 50\% & \bfseries Dice (SD) 25\% & \bfseries Dice (SD) 5\%\\
  \Xhline{2\arrayrulewidth}
  1 & - & DeepLabV3 & 0.9095 (0.0272) & 0.8941 (0.0369) & 0.8731 (0.0646) & 0.7803 (0.1517)\\
  \hline
  2 & ImageNet & DeepLabV3 & 0.9286 (0.0250) & 0.9217 (0.0420) & 0.9120 (0.0456) & 0.8539 (0.0930)\\
  \hline
  3 & SimCLR (C) & DeepLabV3 &  0.9105 (0.0509)  & 0.8862 (0.0606) &  0.8851 (0.0569) & 0.8450 (0.1109)\\
  \hline
  4 & SimCLR (E) & DeepLabV3 & \textbf{0.9311} (0.0424) & 0.9219 (0.0543) & 0.9234 (0.0554)  & \textbf{0.9123} (0.0490)  \\
  \hline
  5 & BYOL (C) & DeepLabV3 & 0.8189 (0.0777) & 0.6202 (0.1275) & 0.5727 (0.1263) & 0.0084 (0.0269)\\
  \hline
  6 & BYOL (E) & DeepLabV3 & 0.8347 (0.1303) & 0.7552 (0.1318) & 0.6321 (0.1363) & 0.5729 (0.2316)\\ 
  \hline
  7 & - & UNet & 0.9125 (0.0263) & 0.8921 (0.0415) & 0.8883 (0.0506)  & 0.8006 (0.1864)\\
  \hline
  8 & SimCLR (C) & UNet & 0.9102 (0.0498) & 0.8965 (0.0663) & 0.8597 (0.1031) & 0.8013 (0.1102)\\
  \hline
  9 & SimCLR (E) & UNet & 0.9296 (0.0506)  & \textbf{0.9224} (0.0798)  & \textbf{0.9248} (0.0596)  & 0.9077 (0.0574) \\
  \hline
  10 & BYOL (C) & UNet & 0.8162 (0.1304) & 0.7810 (0.1263) & 0.7063 (0.1417)  & 0.0520 (0.0907)\\
  \hline     
  11 & BYOL (E) & UNet & 0.8824 (0.0746) & 0.8366 (0.1073) & 0.7984 (0.1418) & 0.7256 (0.1720)\\
  \hline
  \end{tabular}
  \end{adjustbox}}
  \vspace{2mm}*\textit{\textbf{(E):} Pretrained on EchoNet-Dynamic data, \textbf{(C):} Pretrained on CAMUS data}
\end{table}

\section{Results}

\begin{figure}[p]
    \centering
    \null \hspace{4mm} Easy Case \hspace{27mm} Difficult Case \null\\~\\
    \includegraphics[scale=0.6]{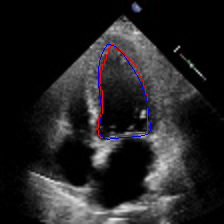}
    \includegraphics[scale=0.6]{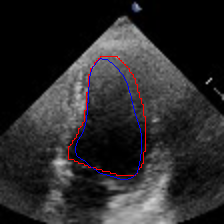} \\[0.05in]
    \includegraphics[scale=0.6]{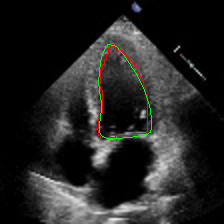}
    \includegraphics[scale=0.6]{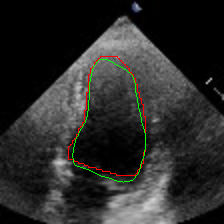}
    \includegraphics[scale=0.6]{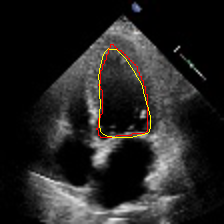}
    \includegraphics[scale=0.6]{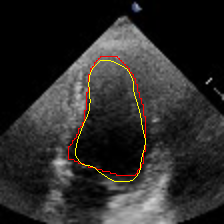}
    \includegraphics[scale=0.6]{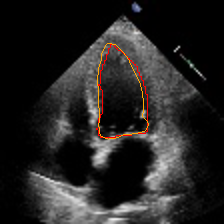}
    \includegraphics[scale=0.6]{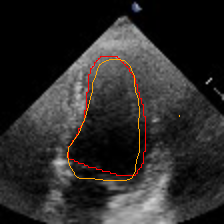}
    \caption{Qualitative results using an ImageNet backbone with 100\% data (blue), a SimCLR backbone with 100\% data (green), a SimCLR backbone with 5\% data (yellow) with DeepLabV3 and a BYOL backbone with 100\% data (orange) with DeepLabV3 on the EchoNet-Dynamic dataset for two cases. \textbf{Right:} Case where predictions are close to the ground truth (red). \textbf{Left:} More difficult case where predictions have more discrepancy.}
    \label{fig:qual_echonet}
\end{figure}

Tables \ref{tab:echo} and \ref{tab:camus} show the quantitative results of the experiments that were conducted, for the EchoNet-Dynamic and CAMUS datasets respectively. Qualitative results from selected models are also shown in Figure \ref{fig:qual_echonet}.

\subsection{EchoNet-Dynamic}

As Table \ref{tab:echo} shows, our proposed self supervised method which uses DeepLabV3 with a SimCLR pretrained backbone outperformed all other methods (including the EchoNet-Dynamic \cite{echonetdynamic} baseline - 0.9211 Dice), regardless of the amount of data. In fact, with only 5\% of the data, our method produces results (0.9125 Dice) that are close to fully supervised training with all of the available data (0.9229 Dice). Furthermore, with the UNet architecture, SimCLR was found to be beneficial although the improvement was minor. We also found ImageNet pretraining to perform better than random initialization. However, BYOL did not have any significant benefit over ImageNet pretraining or even random initialization. Finally, DeepLabV3 performed better than UNet in the segmentation task.

\subsection{CAMUS}

Results on the CAMUS dataset are shown in Table \ref{tab:camus}. When pretrained on CAMUS, SimCLR backbones (0.9105 Dice) were found to perform worse than ImageNet pretrained backbones (0.9286 Dice). However, SimCLR backbones pretrained on the EchoNet-Dynamic dataset showed better performance (0.9311 Dice), exceeding both random initialization and ImageNet pretrained backbones. This was the case for both DeepLabV3 and UNet. Meanwhile, BYOL backbones continued to show worse performance on the CAMUS dataset as well, especially when pretrained on the CAMUS dataset itself. When finetuned on only 5\% of the data, these backbones showed extremely poor performance, failing the downstream segmentation task. Pretraining with EchoNet-Dynamic improved the BYOL backbones, which achieved a Dice score of up to 0.7256 when finetuned on 5\% of the data and up to 0.8824 when finetuned on 100\% of the data. 

\section{Discussion}

The experiments have shown that our proposed self supervised method with SimCLR outperforms the same method which uses BYOL when it comes to pretraining backbones for left ventricle echocardiography segmentation. We also noticed that BYOL is less stable than SimCLR. However, the purpose of the experiments was to study the use of these models with minimal changes and see how they perform out-of-the-box, without extensive tuning. This may be part of the reason why BYOL has shown suboptimal performance.
 
The main difference between the two self supervised frameworks is the fact that BYOL only uses positive pairs, trying to maximize agreement between two augmented versions of a single image (positive pair) and hence find a common representation for them. Conversely, SimCLR tries to maximize agreement between the differently augmented versions of an image (positive pair) while also maximizing disagreement between that image and augmented versions of other images (negative pairs). Meanwhile, BYOL's contrastive learning is implicit and indirectly dependent on the differences between the original images. In our experiments, we use a random frame from each video to introduce some dissimilarity between the original images, however it seems like this is not as effective as the transformations that SimCLR uses.
 
Furthermore, constrastive learning requires large amounts of data to produce good results, which is why pretraining on the CAMUS dataset with only 400 samples was not beneficial (rows 3 \& 6 in Table \ref{tab:camus}). Hence, it makes sense for EchoNet-Dynamic pretraining to be more beneficial and its ability to work with a different dataset shows that generalizable features were learned from the pretraining (rows 4 \& 9 in Table \ref{tab:camus}).
 
Apart from contrastive learning, the experiments also suggest that DeepLabV3 is more effective than UNet for echocardiography segmentation (compare rows 3 \& 6 in Table \ref{tab:echo}). In general, what makes DeepLabV3 perform well is its ASPP module that captures multi-scale representations of high level features extracted by the encoder, making it more resistant to changes in object scales (in this case the size of heart structures), which do vary depending on the heart cycle and the anatomy of the patient's heart.

\section{Conclusion}

While contrastive learning is an open research problem, we conclude from our experiments that vanilla SimCLR pretraining could lead to an improvement in cardiac ultrasound segmentation, especially when annotated data for the downstream task is limited. However, it is crucial to pretrain on a large enough dataset to provide good results. Further experimentation could lead to a better understanding of contrastive learning frameworks in the context of cardiac ultrasound imaging. For example, there is room for improvement on the augmentation strategy used in the SimCLR paper because it targeted natural images not medical images. Optimizing the choice of augmentations might lead to significant improvements and is a good direction for future work in this area. Furthermore, it was out of the scope of this paper to investigate different UNet or DeeplabV3 variations such as nnUNet \cite{isensee2021nnu} and DeeplabV3+ \cite{chen2018encoder} and hence this may be a good research direction to investigate. Additionally, both SimCLR and BYOL are sensitive to batch sizes and require very large batch sizes for optimal performance. Regardless, our work has shown that SimCLR does work with minimal changes and moderate resources.

\bibliographystyle{splncs04}
\bibliography{ref}

\begin{thebibliography}{10}
\providecommand{\url}[1]{\texttt{#1}}
\providecommand{\urlprefix}{URL }
\providecommand{\doi}[1]{https://doi.org/#1}

\bibitem{alsharqi2018artificial}
Alsharqi, M., Woodward, W., Mumith, J., Markham, D., Upton, R., Leeson, P.:
  Artificial intelligence and echocardiography. Echo research and practice
  \textbf{5}(4),  R115--R125 (2018)

\bibitem{ResDUnet}
Amer, A., Ye, X., Janan, F.: Resdunet: A deep learning-based left ventricle
  segmentation method for echocardiography. IEEE Access  \textbf{PP}, ~1--1 (10
  2021). \doi{10.1109/ACCESS.2021.3122256}

\bibitem{chartsias2021contrastive}
Chartsias, A., Gao, S., Mumith, A., Oliveira, J., Bhatia, K., Kainz, B.,
  Beqiri, A.: Contrastive learning for view classification of echocardiograms.
  In: International Workshop on Advances in Simplifying Medical Ultrasound. pp.
  149--158. Springer (2021)

\bibitem{aspp}
Chen, L.C., Papandreou, G., Kokkinos, I., Murphy, K., Yuille, A.: Deeplab:
  Semantic image segmentation with deep convolutional nets, atrous convolution,
  and fully connected crfs. IEEE Transactions on Pattern Analysis and Machine
  Intelligence  \textbf{PP} (06 2016). \doi{10.1109/TPAMI.2017.2699184}

\bibitem{chen2017deeplab}
Chen, L.C., Papandreou, G., Schroff, F., Adam, H.: Rethinking atrous
  convolution for semantic image segmentation. arXiv preprint arXiv:1706.05587
  (2017)

\bibitem{chen2018encoder}
Chen, L.C., Zhu, Y., Papandreou, G., Schroff, F., Adam, H.: Encoder-decoder
  with atrous separable convolution for semantic image segmentation. In:
  Proceedings of the European conference on computer vision (ECCV). pp.
  801--818 (2018)

\bibitem{chen2020simple}
Chen, T., Kornblith, S., Norouzi, M., Hinton, G.: A simple framework for
  contrastive learning of visual representations. In: International conference
  on machine learning. pp. 1597--1607. PMLR (2020)

\bibitem{madgrad}
Defazio, A., Jelassi, S.: Adaptivity without compromise: a momentumized,
  adaptive, dual averaged gradient method for stochastic optimization. arXiv
  preprint arXiv:2101.11075  (2021)

\bibitem{grill2020bootstrap}
Grill, J.B., Strub, F., Altch{\'e}, F., Tallec, C., Richemond, P.H.,
  Buchatskaya, E., Doersch, C., Pires, B.A., Guo, Z.D., Azar, M.G., et~al.:
  Bootstrap your own latent: A new approach to self-supervised learning. arXiv
  preprint arXiv:2006.07733  (2020)

\bibitem{he2016deep}
He, K., Zhang, X., Ren, S., Sun, J.: Deep residual learning for image
  recognition. In: Proceedings of the IEEE conference on computer vision and
  pattern recognition. pp. 770--778 (2016)

\bibitem{isensee2021nnu}
Isensee, F., Jaeger, P.F., Kohl, S.A., Petersen, J., Maier-Hein, K.H.: nnu-net:
  a self-configuring method for deep learning-based biomedical image
  segmentation. Nature methods  \textbf{18}(2),  203--211 (2021)

\bibitem{kusunose2019utilization}
Kusunose, K., Haga, A., Abe, T., Sata, M.: Utilization of artificial
  intelligence in echocardiography. Circulation Journal pp. CJ--19 (2019)

\bibitem{leclerc2019deep}
Leclerc, S., Smistad, E., Pedrosa, J., {\O}stvik, A., Cervenansky, F.,
  Espinosa, F., Espeland, T., Berg, E.A.R., Jodoin, P.M., Grenier, T., et~al.:
  Deep learning for segmentation using an open large-scale dataset in 2d
  echocardiography. IEEE transactions on medical imaging  \textbf{38}(9),
  2198--2210 (2019)

\bibitem{moradi2019mfp}
Moradi, S., Oghli, M.G., Alizadehasl, A., Shiri, I., Oveisi, N., Oveisi, M.,
  Maleki, M., Dhooge, J.: Mfp-unet: A novel deep learning based approach for
  left ventricle segmentation in echocardiography. Physica Medica  \textbf{67},
   58--69 (2019)

\bibitem{ouyang2019echonet}
Ouyang, D., He, B., Ghorbani, A., Lungren, M.P., Ashley, E.A., Liang, D.H.,
  Zou, J.Y.: Echonet-dynamic: a large new cardiac motion video data resource
  for medical machine learning. In: NeurIPS ML4H Workshop: Vancouver, BC,
  Canada (2019)

\bibitem{echonetdynamic}
Ouyang, D., He, B., Ghorbani, A., Lungren, M.P., Ashley, E.A., Liang, D.H.,
  Zou, J.Y.: Echonet-dynamic: a large new cardiac motion video data resource
  for medical machine learning (2019)

\bibitem{ouyang2020video}
Ouyang, D., He, B., Ghorbani, A., Yuan, N., Ebinger, J., Langlotz, C.P.,
  Heidenreich, P.A., Harrington, R.A., Liang, D.H., Ashley, E.A., et~al.:
  Video-based ai for beat-to-beat assessment of cardiac function. Nature
  \textbf{580}(7802),  252--256 (2020)

\bibitem{ronneberger2015u}
Ronneberger, O., Fischer, P., Brox, T.: U-net: Convolutional networks for
  biomedical image segmentation. In: International Conference on Medical image
  computing and computer-assisted intervention. pp. 234--241. Springer (2015)

\bibitem{smistad20172d}
Smistad, E., {\O}stvik, A., et~al.: 2d left ventricle segmentation using deep
  learning. In: 2017 IEEE international ultrasonics symposium (IUS). pp.~1--4.
  IEEE (2017)

\end{thebibliography}
\end{document}